\documentclass[%
 preprint,
 amsmath,amssymb,
 aps,
]{revtex4-2}

\usepackage{graphicx}% Include figure files
\usepackage{dcolumn}% Align table columns on decimal point
\usepackage{bm}% bold math
\usepackage{enumitem}

\begin{document}

% \preprint{Submitted to Physical Review Fluids}

\title{Theoretical framework to surpass the Betz limit \\using unsteady fluid mechanics}% Force line breaks with \\

\author{John O. Dabiri}
%  \altaffiliation[Also at ]{Physics Department, XYZ University.}%Lines break automatically or can be forced with \\
%  \email{Second.Author@institution.edu}
\affiliation{%
 Graduate Aerospace Laboratories (GALCIT) and Mechanical Engineering, California Institute of Technology\\
 Pasadena, California, USA 91125
}%

% \collaboration{MUSO Collaboration}%\noaffiliation

%  \homepage{http://www.Second.institution.edu/~Charlie.Author}
% \affiliation{
%  Second institution and/or address\\
%  This line break forced% with \\
% }%
% \affiliation{
%  Third institution, the second for Charlie Author
% \affiliation{%
%  Authors' institution and/or address\\
%  This line break forced with \textbackslash\textbackslash
% }%

% \collaboration{CLEO Collaboration}%\noaffiliation


\begin{abstract}
The Betz limit expresses the maximum proportion of the kinetic energy flux incident on an energy conversion device that can be extracted from an unbounded flow. The derivation of the Betz limit requires an assumption of steady flow through a notional actuator disk that is stationary in the streamwise direction. The present derivation relaxes the assumptions of steady flow and streamwise actuator disk stationarity, which expands the physically realizable parameter space of flow conditions upstream and downstream of the actuator disk. A key consequence of this generalization is the existence of unsteady motions that can, in principle, lead to energy conversion efficiencies that exceed the Betz limit not only transiently, but also in time-averaged performance. Potential physical implementations of those unsteady motions are speculated.
% \begin{description}
% Secondary publications and information retrieval purposes.
% You may use the \texttt{description} environment to structure your abstract;
% use the optional argument of the \verb+\item+ command to give the category of each item. 
% \end{description}
\end{abstract}

\maketitle

%\tableofcontents

The Betz limit~\cite{Betz1926} expresses the currently accepted theoretical limit on the power conversion efficiency of fluid dynamic energy harvesting devices operating in unbounded flow. Indeed, modern wind and hydrokinetic energy conversion devices in nominally unbounded flow exhibit efficiencies below the Betz limit, tacitly supporting its veracity~\cite{Hau2005}. Fundamental to both the Betz limit and the design of typical fluid dynamic energy conversion devices is an assumption that the flow is nominally steady. This steady flow assumption inextricably links the pressure and the velocity along streamlines upstream and downstream of the energy conversion device. Consequently, an unavoidable trade-off exists between the pressure drop that can be induced by the actuator disk and the mass flux through it. Their combination determines the power that can be extracted by the energy conversion device. Betz~\cite{Betz1926} showed that the steady flow trade-off is optimized at a power conversion efficiency of 16/27 or 59.3\%.

While the assumption of steady flow simplifies the fluid dynamic analysis, a much larger parameter space of pressure and velocity is accessible if we relax the requirement of steady flow, and we instead permit unsteady streamwise motion of the actuator disk. Previous work (e.g.~\cite{Chattot2014}) has suggested that the introduction of unsteady fluid mechanics at the actuator disk can transiently increase the power conversion efficiency above the Betz limit. However, the time-averaged performance in those cases has still remained bounded by the steady flow limit. Furthermore, those results have not explicitly accounted for the energy required to generate the unsteady actuator disk motion. In the treatment that follows, we include a full accounting of the energy cost of unsteadiness as we examine the potential to leverage unsteady fluid mechanics to surpass the Betz limit in time-averaged performance.

Consider a flow from left to right through an actuator disk (Fig. 1). The upstream flow station is denoted 1. The locations immediately upstream and downstream from the actuator disk are denoted 2 and 3, respectively. The flow far downstream is denoted station 4.

\begin{figure}
\centering
\includegraphics[width=5in]{{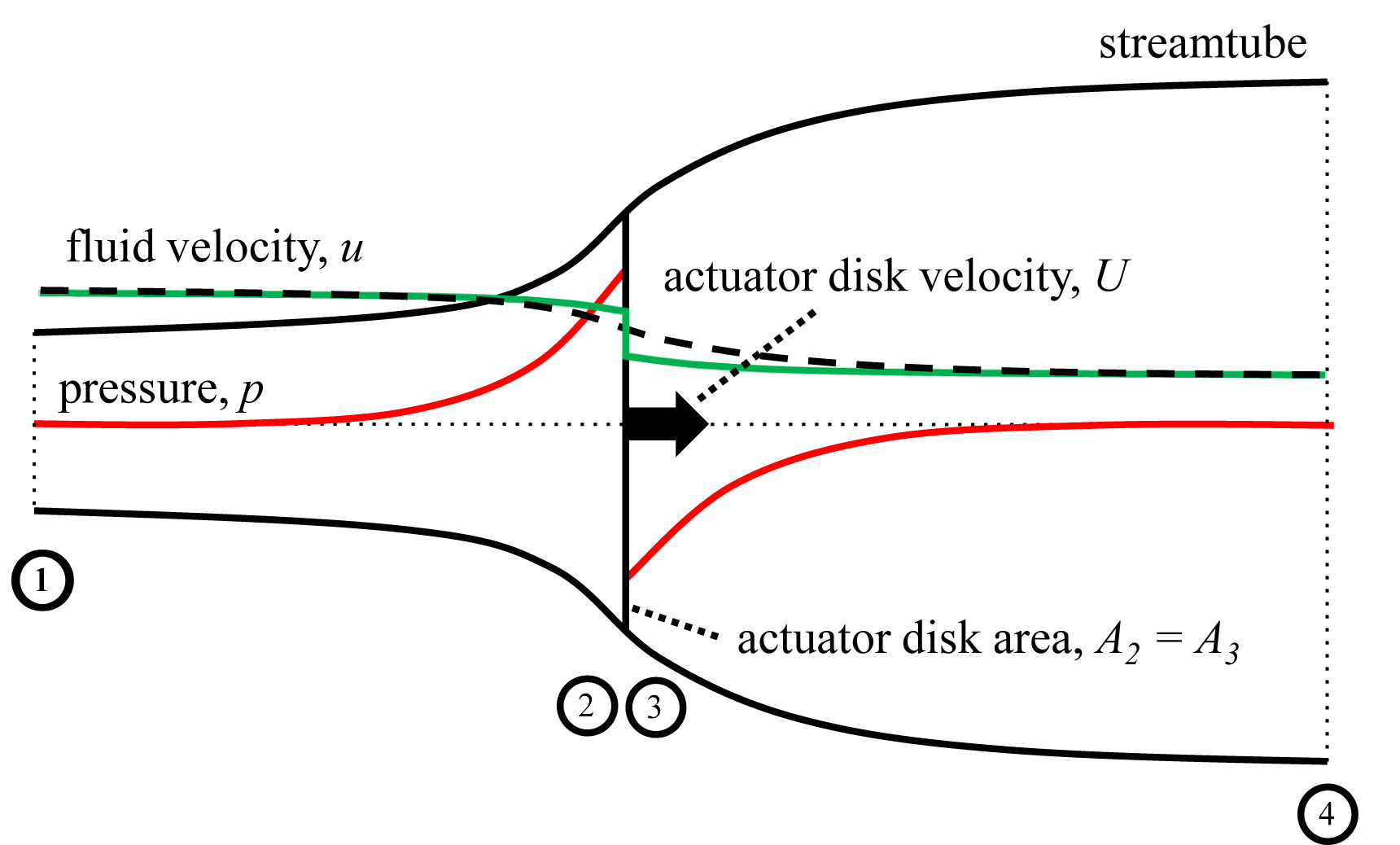}} 
\caption{Schematic of flow through a fluid dynamic energy conversion device, idealized as an actuator disk located between stations 2 and 3. Qualitative profiles of steady flow pressure (red) and velocity (black dashed) are illustrated. The modified velocity profile due to unsteady flow is illustrated by the solid green curve. Adapted from~\protect\cite{Araya_etal_2014}.  }
\end{figure}

Assuming inviscid, unsteady flow between stations 1 and 2, the flow along an unsteady streamline connecting these stations is given by

\begin{equation} \label{unsteadybernoulli12}
p_2 = p_1 + \frac{1}{2} \rho \big( u_1^2 - u_2^2 \big) + \rho\frac{\partial}{\partial t}\big(\phi_1 - \phi_2\big)
\end{equation}

\noindent where $\rho$ is the fluid density, and $p_i$, $u_i$, and $\phi_i$ are the pressure, flow speed, and velocity potential, respectively, at station $i$. Similarly, the flow properties at stations 3 and 4 are related as

\begin{equation} \label{unsteadybernoulli34}
p_3 = p_4 + \frac{1}{2} \rho \big( u_4^2 - u_3^2 \big) + \rho\frac{\partial}{\partial t}\big(\phi_4 - \phi_3\big)
\end{equation}

The change in momentum flux from station 1 to station 4 due to the presence of the actuator disk is given by the force of the actuator disk on the flow:

\begin{equation} \label{thrusteqn}
\rho u_4^2 A_4 - \rho u_1^2 A_1 = -\big(p_2-p_3\big) A_2
\end{equation}

\noindent where $A_i$ is the cross sectional area at station $i$ of the smallest streamsurface that encompasses the actuator disk. 

Finally, a portion of the kinetic energy incident on the actuator disk at station 2 can be used to create unsteady motion of the fluid surrounding the actuator disk. The energy of that unsteady fluid motion is

\begin{equation} \label{diskenergy}
KE_{disk} = -\frac{1}{2} \rho \oint \phi \mathbf{n} \cdot \nabla \phi \,dA = \rho \Phi_s U
\end{equation}

\noindent where $\Phi_s$ is the component of the unsteady velocity potential in the direction of the streamwise unit vector $\hat{\mathbf{i}}$, i.e. $\Phi_s = \hat{\mathbf{i}} \cdot -\frac{1}{2}\oint \phi \mathbf{n} \,dA$, and $U$ is the streamwise component of the actuator disk velocity in a lab frame, i.e. $U = \hat{\mathbf{i}} \cdot \nabla \phi$~\cite{Batchelor2000}.

Substituting equations~\ref{unsteadybernoulli12} and~\ref{unsteadybernoulli34} for $p_2$ and $p_3$ in equation~\ref{thrusteqn}:

\begin{multline} \label{modifiedthrusteqn}
    \rho u_4^2 A_4 - \rho u_1^2 A_1 = -\Big(p_1 + \frac{1}{2} \rho \big( u_1^2 - u_2^2 \big) + \rho\frac{\partial}{\partial t}\big(\phi_1 - \phi_2\big)\\-p_4 - \frac{1}{2} \rho \big( u_4^2 - u_3^2 \big) - \rho\frac{\partial}{\partial t}\big(\phi_4 - \phi_3\big)\Big) A_2
\end{multline} 

Let us assume that the pressure at stations 1 and 4 is atmospheric. With this assumption, equation~\ref{modifiedthrusteqn} becomes

\begin{multline} \label{simplifiedthrusteqn1}
    u_4^2 A_4 - u_1^2 A_1 = \Big(-\frac{1}{2} u_1^2 +\frac{1}{2} u_2^2 -\frac{1}{2} u_3^2 +\frac{1}{2} u_4^2  - \frac{\partial \phi_1}{\partial t} + \frac{\partial \phi_2}{\partial t} - \frac{\partial \phi_3}{\partial t} + \frac{\partial \phi_4}{\partial t} \Big) A_2
\end{multline} 

\noindent or

\begin{equation} \label{simplifiedthrusteqn2}
    u_4^2 A_4 - u_1^2 A_1 = \Big(-\frac{1}{2} u_1^2 +\frac{1}{2} u_2^2 -\frac{1}{2} u_3^2 +\frac{1}{2} u_4^2  + \Phi_t \Big) A_2
\end{equation} 

\noindent where $\Phi_t$ henceforth captures the unsteady terms arising from the streamwise actuator disk motion, i.e. $\frac{\partial \phi_2}{\partial t} - \frac{\partial \phi_3}{\partial t}$.

By conservation of mass, $u_1 A_1 = u_2 A_2 = u_4 A_4$. Therefore, $A_1$ and $A_4$ in equation~\ref{simplifiedthrusteqn2} can be replaced by $A_2$ as

\begin{equation} \label{simplifiedthrusteqnwithA2_1}
    \frac{u_4^2 u_2 A_2}{u_4} - \frac{u_1^2 u_2 A_2}{u_1} = \Big(-\frac{1}{2} u_1^2 +\frac{1}{2} u_2^2 -\frac{1}{2} u_3^2 + \frac{1}{2} u_4^2 + \Phi_t \Big) A_2
\end{equation}

\begin{equation} \label{simplifiedthrusteqnwithA2_2}
    u_4 u_2 - u_1 u_2 = -\frac{1}{2} u_1^2 +\frac{1}{2} u_2^2 -\frac{1}{2} u_3^2 + \frac{1}{2} u_4^2 + \Phi_t 
\end{equation}

\begin{equation} \label{simplifiedthrusteqnwithA2_3}
    u_2\big(u_4  - u_1\big) = \frac{1}{2}\big(u_4 + u_1\big) \big(u_4 - u_1\big) + \frac{1}{2}\big(u_2 + u_3\big) \big(u_2 - u_3\big) + \Phi_t 
\end{equation}

From equation~\ref{diskenergy}, balancing the kinetic energy flux across the actuator disk and the kinetic energy associated with streamwise actuator disk motion, 

\begin{equation} \label{diskenergy2}
    \big(u_2^3 - u_3^3\big)A_2 = 2\Big(U\frac{d\Phi_s}{dt} + \frac{dU}{dt}\Phi_s\Big) \approx 2U\frac{d\Phi_s}{dt} 
\end{equation}

\noindent where the approximation assumes that temporal variation is dominated by the unsteady velocity potential. We will revisit this assumption in the concluding discussion. Equation~\ref{diskenergy2} can be further simplified as

\begin{equation} \label{diskenergy3}
    \big(u_2 - u_3\big)\big(u_2^2 +u_2 u_3 + u_3^2 \big)A_2  = 2U\frac{d\Phi_s}{dt} 
\end{equation}

\begin{equation} \label{diskenergy4}
    \big(u_2 - u_3\big)\Big(\big(u_2 + u_3\big)^2 - u_2 u_3\Big)A_2  = 2U\frac{d\Phi_s}{dt}  
\end{equation}

Now, let us approximate the time derivative of $\Phi_s$ as

\begin{equation} \label{Phis_Phit}
    \frac{d\Phi_s}{dt} \approx A_2\Phi_t
\end{equation}

\noindent which effectively assumes that $\phi$ does not exhibit substantial spatial variability at the actuator disk, and that the actuator disk area does not exhibit substantial temporal variability (assumptions that will also be revisited in the concluding discussion).

Substituting equation~\ref{Phis_Phit} into~\ref{diskenergy4} and solving for $\Phi_t$ gives

\begin{equation} \label{phi_t_eqn}
    \Phi_t = \frac{1}{2U}\big(u_2 - u_3\big)\Big(\big(u_2 + u_3\big)^2 - u_2 u_3\Big) 
\end{equation}

Replacing $\Phi_t$ in equation~\ref{simplifiedthrusteqnwithA2_3} with equation~\ref{phi_t_eqn} gives

\begin{multline} \label{simplifiedthrusteqnwithA2_4}
    u_2\big(u_4  - u_1\big) = \frac{1}{2}\big(u_4 + u_1\big) \big(u_4 - u_1\big) + \frac{1}{2}\big(u_2 + u_3\big) \big(u_2 - u_3\big)\\ + \frac{1}{2U}\big(u_2 - u_3\big)\Big(\big(u_2 + u_3\big)^2 - u_2 u_3\Big) 
\end{multline}

Note that in the case of steady flow, $\Phi_t = 0$, $u_2 = u_3$ (cf. equation~\ref{phi_t_eqn} or setting $A_2 = A_3$ in the continuity equation), and we recover the result from the classical Betz derivation that the velocity at the actuator disk is the average of the upstream and downstream flow speeds, i.e. $u_2 = u_3 = \frac{1}{2}\big(u_1+u_4)$.

The power extracted by the actuator disk is given by the product of the mass flux through the actuator disk and the difference in kinetic energy upstream and downstream of the actuator disk:

\begin{equation} \label{Peqn}
P = \frac{1}{2}\rho A_2 u_2\big(u_1^2 - u_4^2\big) = \frac{1}{2}\rho A_2 u_2\big(u_1 - u_4\big)\big(u_1 + u_4\big)
\end{equation}

Note that this extracted power comprises steady and unsteady components, both of which are assumed to be convertible to useful work. Substituting for $u_2$ from equation~\ref{simplifiedthrusteqnwithA2_4} gives

\begin{multline} \label{simplifiedPeqn}
P = \frac{1}{2}\rho A_2 \Bigg[\frac{1}{2}\big(u_1 + u_4\big)^2 \big(u_1 - u_4\big) - \frac{1}{2}\big(u_1 + u_4\big)\big(u_2 + u_3\big) \big(u_2 - u_3\big)\\ + \frac{\big(u_1 + u_4\big)\big(u_2 - u_3\big)}{2U}\Big(u_2 u_3 - \big(u_2 + u_3\big)^2\Big)\Bigg] 
\end{multline}

Define a power coefficient, $C_p \equiv P/\big(\frac{1}{2}\rho A_2 u_1^3\big)$, which quantifies the efficiency of fluid dynamic energy conversion. Substituting for $P$ from equation~\ref{simplifiedPeqn} gives

\begin{multline} \label{simplifiedCpeqn}
C_p = \frac{\big( u_1 + u_4\big)^2\big( u_1 - u_4\big)}{2u_1^3} - \frac{\big(u_1 + u_4\big)\big(u_2 + u_3\big) \big(u_2 - u_3\big)}{2u_1^3}\\ + \frac{\big(u_1 + u_4\big)\big(u_2 - u_3\big)}{2U u_1^3}\Big(u_2 u_3 - \big(u_2 + u_3\big)^2\Big)
\end{multline}

Equation~\ref{simplifiedCpeqn} can be rewritten in terms of the ratios $u_2/u_1$, $u_3/u_1$, and $u_4/u_1$ as

\begin{multline} \label{simplifiedCpeqn_ratios}
C_p = \frac{1}{2}\Bigg[1+\frac{u_4}{u_1}-\Big(\frac{u_4}{u_1}\Big)^2 -\Big(\frac{u_4}{u_1}\Big)^3\Bigg]\\ + \frac{1}{2}\Bigg[-\Big(\frac{u_2}{u_1}\Big)^2 -\Big(\frac{u_4}{u_1}\Big)\Big(\frac{u_2}{u_1}\Big)^2 + \Big(\frac{u_3}{u_1}\Big)^2 + \Big(\frac{u_4}{u_1}\Big)\Big(\frac{u_3}{u_1}\Big)^2\Bigg]\\
+ \frac{1}{2}\Bigg[-\Big(\frac{u_1}{U}\Big)\Big(\frac{u_2}{u_1}\Big)^3 + \Big(\frac{u_1}{U}\Big)\Big(\frac{u_3}{u_1}\Big)^3 - \Big(\frac{u_4}{U}\Big)\Big(\frac{u_2}{u_1}\Big)^3 +\Big(\frac{u_4}{U}\Big)\Big(\frac{u_3}{u_1}\Big)^3\Bigg]
\end{multline}

Now, let us define an \textit{actuator induction coefficient} $a \equiv \big(u_1 - u_2\big)/u_1$; a \textit{near-wake induction coefficient} $b \equiv \big(u_1 - u_3\big)/u_1$; and a \textit{far-wake induction coefficient} $c \equiv \big(u_1 - u_4\big)/u_1$. Reformatting equation~\ref{simplifiedCpeqn_ratios} in terms of these coefficients gives

\begin{multline} \label{simplifiedCpeqn_coeffs}
C_p = \frac{1}{2}\big(4c - 4c^2 + c^3\big) + \frac{1}{2}\Big\{\big(2-c\big)\Big[\big(1-b\big)^2-\big(1-a\big)^2\Big]\Big\}\\
+ \frac{1}{2}\Big\{\Big(\frac{u_1}{U}\Big)\Big(2-c\Big)\Big[\big(1-b\big)^3-\big(1-a\big)^3\Big]\Big\}
\end{multline}

The steady flow assumption of Betz~\cite{Betz1926} posits that the flow speed immediately upstream and downstream of the actuator disk is identical, i.e. $a = b$. Hence, only the first term in equation~\ref{simplifiedCpeqn_coeffs} remains. The necessary condition for maximum efficiency in that case is

\begin{equation} \label{maxeffeqn}
\frac{\partial C_p}{\partial c} = \frac{1}{2}\big(4 - 8c + 3c^2\big) = 0
\end{equation}

\noindent or

\begin{equation} \label{maxeffeqn_2}
\big(2-c\big)\big(2-3c\big) = 0
\end{equation}

Since the only physical solutions are $0 \leq c \leq 1$, i.e. the far-wake speed is non-negative and no greater than the upstream flow speed, the only physical root of equation~\ref{maxeffeqn_2} is $c = 2/3$. In other words, efficiency in the steady flow case is maximized when the far-wake flow speed is reduced by $2/3$ from the upstream flow speed. From equation~\ref{simplifiedCpeqn_coeffs} with $a = b = c/2$ (per the steady flow limit of equation~\ref{simplifiedthrusteqnwithA2_4}), the corresponding efficiency in this case is $C_p(c=2/3) = 16/27$ or 59.3\%. This is known as the Betz limit.

In the more general case of unsteady flow, $b$ can be greater than $a$ (i.e. $u_3$ less than $u_2$) with their difference contributing to the energy of unsteady motion $\rho\Phi_s U$ at the actuator disk. That motion can in turn be leveraged to modify the power coefficient, by decoupling the pressure and velocity in equations~\ref{unsteadybernoulli12} and~\ref{unsteadybernoulli34} (i.e. enabling a larger pressure gradient across the actuator disk without compromising the mass flux through the system). Examination of equation~\ref{simplifiedCpeqn_coeffs} shows that if $b > a$, the second term is negative, meaning that the power coefficient is reduced relative to the steady case. The quantity in square brackets in the third term of equation~\ref{simplifiedCpeqn_coeffs} will also be negative, meaning the only way to increase the power coefficient for $b > a$ is if $U$ is negative, i.e. the actuator disk exhibits upstream motion. Note that this upstream motion need not be maintained for all time. Rather, the unsteady motion need only achieve a \textit{weighted, time-averaged} $U$ that is negative, in order for the time-averaged power coefficient to exceed the Betz limit. The pertinent weight for $U$ is the nonlinear function of the induction coefficients $a$, $b$, and $c$ in the third term of equation~\ref{simplifiedCpeqn_coeffs}. Because the instantaneous values of the induction coefficients can vary in time, a negative weighted time-average of $U$ (corresponding to a positive time-averaged value of the third term in equation~\ref{simplifiedCpeqn_coeffs}) can be achieved despite the fact that the time-average of $U$ itself is zero, e.g. if we require that the actuator disk has no net streamwise displacement over time. Candidate physical implementations are described in the concluding discussion.

To summarize, the optimization of power conversion efficiency in the generalized case of unsteady flow is governed by equation~\ref{simplifiedCpeqn_coeffs}, subject to the kinetic energy constraint of equation~\ref{diskenergy2}, i.e. 

\begin{equation} \label{energyconstraint}
\big(1-a\big)^3 - \big(1-b\big)^3 = \frac{2\Phi_t}{u_1^2}\Big(\frac{U}{u_1}\Big)
\end{equation}

\noindent with $\text{sgn}(\Phi_t) = \text{sgn}(U)$ because the kinetic energy in equation~\ref{diskenergy} is non-negative; and $a \leq b \leq c$, because the actuator disk exerts a net drag that decelerates the flow. 

It is important to consider equations~\ref{simplifiedCpeqn_coeffs} and~\ref{energyconstraint} concurrently in order to gain intuition for the effect of flow unsteadiness on the power conversion efficiency. For example, examination of equation~\ref{simplifiedCpeqn_coeffs} reveals a non-trivial dependence of power conversion efficiency on the unsteady actuator disk velocity $U$ in the third term. As described above, this term augments the power coefficient if $U < 0$ and $a < b$. However, because the product of $U$ and $\Phi_t$ is constrained by the relative values of $a$ and $b$ (cf. equation~\ref{energyconstraint}), an increase in $\|U\|$ can require a decrease in $\|\Phi_t\|$, the other essential unsteady parameter. The appearance of the unsteady flow velocity $U$ in the denominator of the third term of equation~\ref{simplifiedCpeqn_coeffs} reflects that trade-off between the effect of increasing unsteady flow magnitude $\|U\|$ and the concomitant decrease in the magnitude of the unsteady velocity potential $\Phi_t$, per equation~\ref{energyconstraint}.

Mathematically, the power coefficient $C_p$ is not bounded by 1 for all $U$ and $\Phi_t$. Physically, however, the power coefficient will be limited by the ability of the incident flow to create the unsteadiness quantified by $U$ and $\Phi_t$.  Given our interest in physically realizable conditions, and in light of the relatively high dimensionality of the parameter space, let us limit our investigation to flow unsteadiness that is commensurate with the incoming flow, i.e. $\| U \| \sim u_1 $ and $\| \Phi_t \| \sim  u_1^2 $. A multi-parameter optimization of the induction coefficients $a$, $b$, and $c$ in the power coefficient equation~\ref{simplifiedCpeqn_coeffs}, constrained by the unsteady actuator disk equation~\ref{energyconstraint}, characterizes the regime in which the Betz limit can be exceeded and to what extent. Using a constrained nonlinear multivariable function minimizer (\textit{fmincon} function in Matlab) initialized using the optimal steady flow parameters (i.e. $a_0 = b_0 = 1/3,\, c_0 = 2/3$, and  $\big(U/u_1\big)_0 = \big(\Phi_t/u_1^2\big)_0 = 0$) one finds that a local maximum in power coefficient exists for $a = 0.0119$, $b = 0.4762$, $c = 0.6236$, $U/u_1 = -0.6804$, and $\Phi_t/u_1^2 = -0.6010$. The corresponding power coefficient for these parameters is $C_p$ = 93.8\%, which significantly exceeds the Betz limit. 

Figure 1 contrasts the trend in flow speed for this scenario versus a steady flow case. The velocity drop across the actuator disk provides the kinetic energy of the unsteady streamwise actuator disk motion. In turn, the unsteady streamwise motion of the actuator disk engenders a time-dependent velocity potential at stations 2 and 3, which enables the pressure drop to be decoupled from the flow velocity. This effect is captured by the non-zero terms $\frac{\partial \phi_2}{\partial t}$ and $\frac{\partial \phi_3}{\partial t}$ that now arise in equations~\ref{unsteadybernoulli12} and~\ref{unsteadybernoulli34}. Because the pressure drop is decoupled from the flow velocity, we have removed the steady flow trade-off between those factors, which together dictate the amount of power that can be extracted from the flow. Notably, in the parameter regime where $b \rightarrow c$, i.e. where the near- and far-wake flow become identical, the power conversion efficiency increases even further beyond the local maximum identified above. However, this result is likely non-physical in the limit, as it corresponds to instantaneous near-wake recovery to the downstream ambient flow conditions. Finally, mass conservation can be ensured by a discontinuous increase in the size of the bounding streamsurface downstream of the actuator disk, or by entrainment of flow from the lateral and vertical directions in the actuator disk wake. Either of these options is less severe than the discontinuous change in mass flux required in previous unsteady models in order to transiently exceed the Betz limit~\cite{Chattot2014}.  

To illustrate a time-dependent variation in $U$ that introduces these unsteady effects but remains zero in time-average (i.e. to avoid net streamwise displacement of the actuator disk over time), figure 2 plots the optimal values of $a$, $b$, $c$, $\Phi_t/u_1^2$, and $C_p$ for a periodic, stepwise variation of $U/u_1$ with zero time-average. The bounds on $U/u_1$ are set at $\pm 0.6804$ to match the local maximum identified above. A corresponding periodic, stepwise variation in the power coefficient is observed, with a maximum value of $C_p = 93.8\%$ and a minimum value of $C_p = 59.0\%$. This minimum value is just below the Betz limit, and it is achieved for values of $a$, $b$, and $c$ similar to the optimal steady flow case. The potentially deleterious effect of positive $U$ on the power coefficient (cf. equation~\ref{simplifiedCpeqn_coeffs}) is mitigated because the concurrent unsteady velocity potential is small, i.e $\Phi_t/u_1^2 = 0.001$, resulting in $a \approx b$ during this motion of the actuator disk.

\begin{figure}
\centering
\includegraphics[width=7in]{{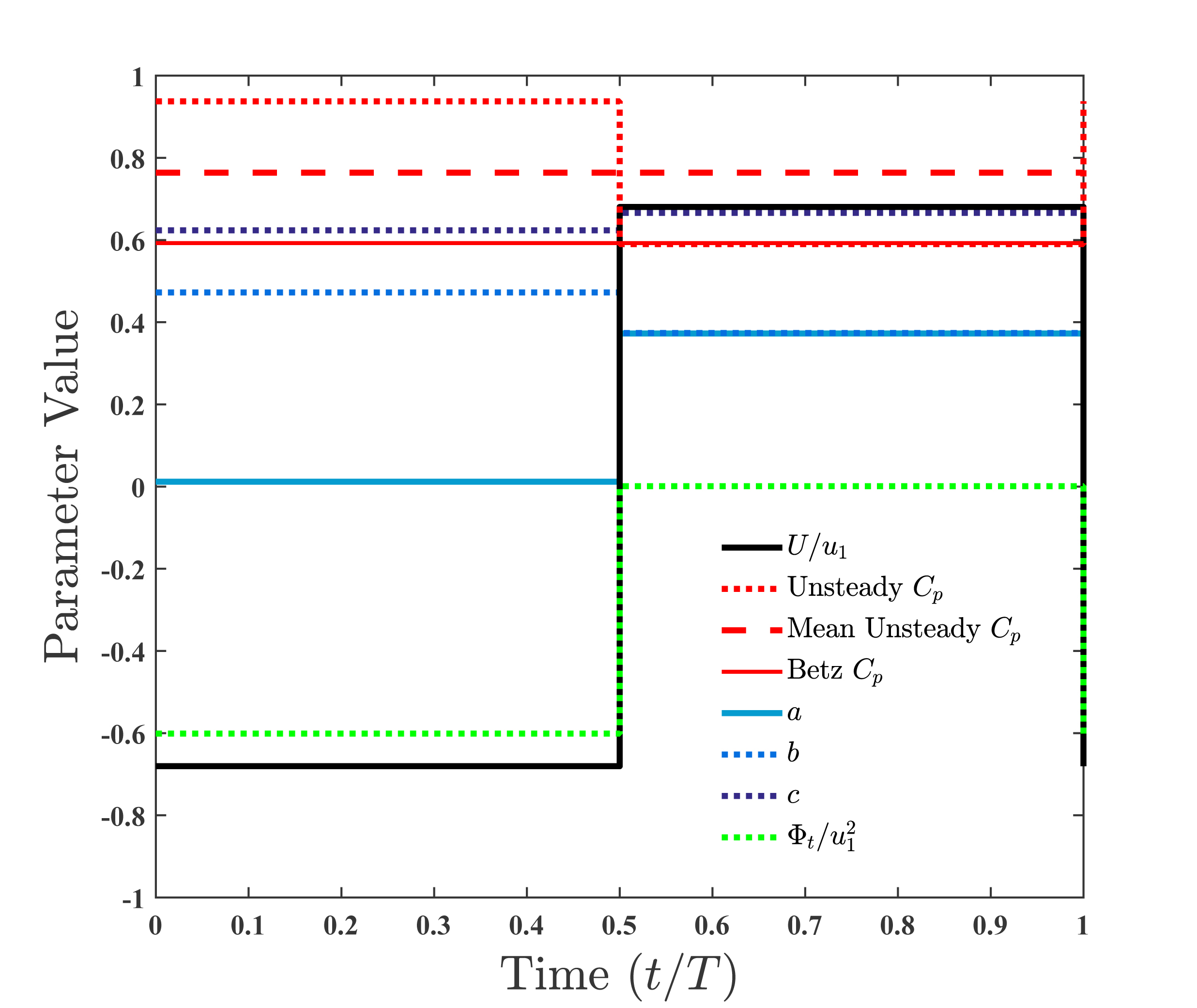}} 
\caption{Power coefficient and flow parameters for unsteady actuator disk over a period $T$ of unsteady motion. The Betz limit is exceeded for an unsteady actuator disk motion with zero time-averaged velocity. Legend inset.}
\end{figure}

The time-averaged power coefficient is $C_p = 76.4\%$, which again exceeds the Betz limit, in spite of the additional constraint placed on the time-average of $U$ to prohibit net streamwise motion of the actuator disk. Higher time-averaged power coefficients can potentially be achieved by optimization of the time-dependent actuator disk motion $U(t)$.

The foregoing derivation leveraged two simplifying assumptions regarding the form of the velocity potential term. First, it was assumed in equation~\ref{diskenergy2} that the temporal variation in the velocity potential of the actuator disk dominates the temporal variation of its streamwise velocity. If we approximate the time derivatives in equation~\ref{diskenergy2} as $\frac{d\Phi_s}{dt} \approx \Delta \Phi_s/\Delta t$ and $\frac{dU}{dt} \approx \Delta U/\Delta t$, respectively, then the assumption used in equation~\ref{diskenergy2} is effectively that $\|\big(\Delta \Phi_s\big) / \Phi_s\| \gg \|\big(\Delta U\big) / U\|$. Since $U$ must be non-zero in the unsteady case (e.g. in the present optimal condition, $U \sim u_1$), the inequality can be satisfied if $\Phi_s$ remains small in magnitude. This occurs, for example, if $\phi$ in equation~\ref{diskenergy} does not vary significantly along the actuator disk surface. Indeed, in the limit of a spatially constant $\phi$, the spatial integral $\Phi_s \rightarrow 0$ by definition. Note that the unsteady flow physics demand a non-zero value of $\frac{d\Phi_s}{dt}$ to realize the efficiency benefits described herein. However, no constraint exists on $\Phi_s$ instantaneously; it can oscillate about zero to satisfy the approximation in equation~\ref{diskenergy2} without diminishing the unsteady fluid mechanics.

The approximation in equation~\ref{Phis_Phit} likewise assumes that the velocity potential $\phi$ does not vary significantly across the actuator disk, so that the velocity potential can be pulled out of the spatial integral in equation~\ref{diskenergy} before differentiation in time. If the cross-sectional area of the actuator disk is also approximately constant, then its time-derivative vanishes in application of Leibniz's rule to the integral, resulting in the approximation in equation~\ref{Phis_Phit}.

The aforementioned assumptions can potentially be satisfied by an energy conversion device comprising rigid structures (e.g. airfoils) that exhibit a component of rotational motion in a plane parallel to the streamwise direction. The angular motion of the structures changes their shape and orientation relative the the streamwise direction, with a corresponding time-dependent variation of $\Phi_s$. In addition, if the actuator disk motion is reciprocal, fore-aft shape asymmetry of the structures can facilitate a larger, non-zero value of $\Phi_t$ during upstream motion and a value of $\Phi_t$ approaching zero during downstream motion, mimicking the solution in figure 2. Finally, if the axis of rotation is fixed, then the rigid, rotating structures will also maintain a constant, time-averaged cross-sectional area. Excluding the reciprocal motion, such a design shares some conceptual similarities to the Pelton wheel~\cite{Sayers1990}, which converts the kinetic energy of an impinging liquid stream in air into useful work. The Pelton wheel also exhibits power conversion efficiencies that well exceed the Betz limit~\cite{SolemslieDahlhaug_2012}; however, the trade-off between pressure drop and mass flux is inherently avoided in that system, because the incident liquid stream is accumulated in a separate discharge reservoir. No such mass reservoir is available in the present context of single-phase fluid dynamic energy conversion devices. Instead, the preceding analysis suggests that it may be possible to surpass the Betz limit by exploiting unsteady fluid mechanics. The present theoretical framework can potentially guide the design, characterization, and optimization of such unsteady fluid dynamic energy conversion devices.

\begin{acknowledgments}
The author gratefully acknowledges helpful feedback from Robert Whittlesey and Daniel Araya.
\end{acknowledgments}

\bibliography{apssamp}% Produces the bibliography via BibTeX.

\end{document}